\documentclass[11pt]{iopart}
\usepackage{graphicx}
\usepackage{xcolor}
\usepackage{witharrows}
\expandafter\let\csname equation*\endcsname\relax
\expandafter\let\csname endequation*\endcsname\relax
\usepackage{amsmath}
\usepackage{physics}
\usepackage[bookmarks=true, hidelinks]{hyperref}
\usepackage[version=4]{mhchem}
\usepackage{booktabs}
\usepackage[color=cyan, disable=true]{todonotes}
\usepackage{svg}
\newcommand{\be}{\begin{equation}}
\newcommand{\ee}{\end{equation}}
\usepackage[style=nature, isbn=false, date=year]{biblatex}
\begin{document}
\title[Complex quantum network models from spin clusters]{Complex quantum network models from spin clusters}
\author{Ravi T. C. Chepuri$^{1,2}$ and Istv\'an A. Kov\'acs$^{1,3}$}
\address{$^1$ Department of Physics and Astronomy, Northwestern University, Evanston, IL 60208, USA}
\address{$^2$ Department of Physics, University of Maryland, College Park, MD 20742, USA}
\address{$^3$ Northwestern Institute on Complex Systems, Northwestern University, Evanston, IL 60208, USA}
\ead{istvan.kovacs@northwestern.edu}

%%%%%%%%%%%%%%%%
% ABSTRACT
%%%%%%%%%%%%%%%%

\begin{abstract}

In the emerging quantum internet, complex network topology could lead to efficient quantum communication and enhanced robustness against failures. However, there are some concerns about complexity in quantum communication networks, such as potentially limited end-to-end transmission capacity. These challenges call for model systems in which the feasibility and impact of complex network topology on quantum communication protocols can be explored. Here, we present a theoretical model for complex quantum communication networks on a lattice of spins, wherein entangled spin clusters in interacting quantum spin systems serve as communication links between appropriately selected regions of spins. Specifically, we show that ground state Greenberger–Horne–Zeilinger clusters of the two-dimensional random transverse Ising model can be used as communication links between regions of spins, and we show that the resulting quantum networks can have complexity comparable to that of the classical internet. Our work provides an accessible generative model for further studies towards determining the network characteristics of the emerging quantum internet.

\end{abstract}

\maketitle

%%%%%%%%%%%%%%%%
% INTRODUCTION
%%%%%%%%%%%%%%%%

\section{Introduction}

The creation of a global quantum communication network to serve as a quantum internet is a highly anticipated goal that would enhance the existing classical internet, with applications in secure communications, quantum computation, distributed sensing, and more \cite{kimble_2008, wehner.etal_2018, kozlowski.wehner_2019, cacciapuoti.etal_2020}. Experimental advances toward the technology needed for a quantum internet, such as quantum repeaters, are rapidly being made  \cite{vanleent.etal_2022, luo.etal_2022, wei.etal_2022, chen.etal_2020, fang.etal_2020, yin.etal_2017, liao.etal_2017, ren.etal_2017}. Such advances have already enabled the creation of quantum communication networks with a few nodes, which may be precursors to the future quantum internet  \cite{peev.etal_2009, sasaki.etal_2011, liao.etal_2018, wehner.etal_2018}.

\begin{figure}[t]
	\begin{center}
	    \includegraphics[width=1\textwidth]{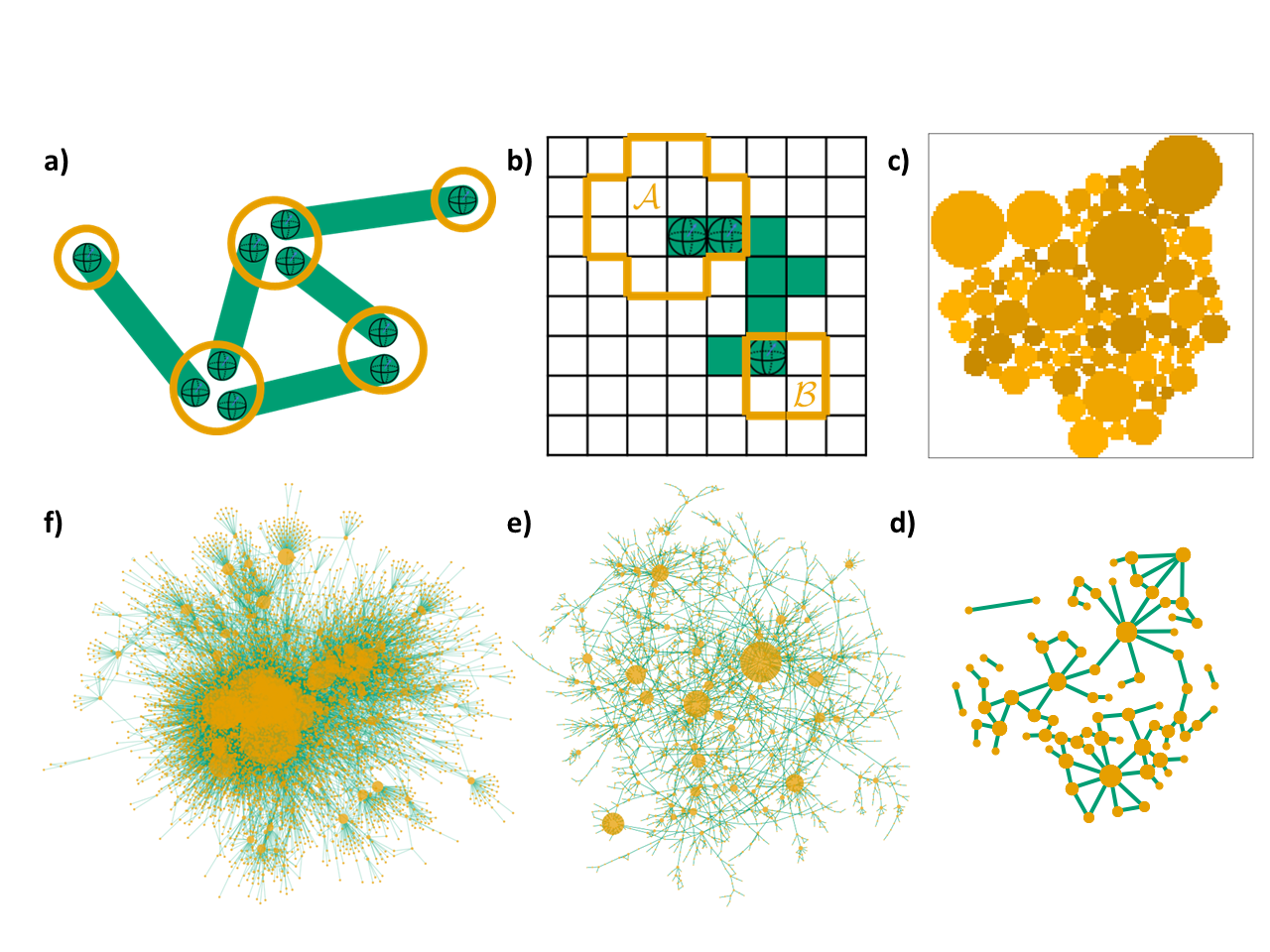}
	    \caption{\textbf{Creation of quantum networks using spin clusters.}
        	    \textbf{a)} Quantum communication networks based on quantum repeaters consist of collections of local qubits (orange nodes) that are entangled (green links) with qubits in other local collections.
        	    \textbf{b)} We propose the use of entangled clusters (green) in interacting spin systems as communication channels between nodes that are spatially localized regions of lattice sites (orange).
        	    \textbf{c)} Nodes are chosen as adjacent circular regions of sites from a broad size distribution, as illustrated here on a lattice of size $128 \times 128$. A link is added when a cluster in the underlying spin system is shared by exactly two nodes. 
        	    \textbf{d)} The resulting network between nodes in panel c) indicates which pairs of nodes share entanglement.
        	    \textbf{e)} Using larger lattices (here $4096 \times 4096$), a complex network emerges.
        	    \textbf{f)} A snapshot of the classical internet from 1999 at the level of autonomous systems \cite{meyer_} shows similar complexity to that in panel e), although being significantly larger.
        	    }
	    \label{fig:panel_1}
	\end{center}
\end{figure}

In the simplest case of quantum repeater based architectures, quantum communication networks can be represented as shown in Figure \ref{fig:panel_1}a \cite{biamonte.etal_2019, acin.etal_2007, cirac.etal_1997, satoh.etal_2012, perseguers.etal_2008, schoute.etal_2016, meignant.etal_2019}. Each node of the network represents a collection of quantum bits, or qubits, and each link of the network represents an entangled pair of qubits belonging to two distinct nodes. As such, the degree of each node (the number of links it has) is at most the number of qubits in the node. Quantum routing protocols based on entanglement swapping can generate entanglement between nodes which were initially not adjacent in the network, thereby enabling quantum communication between remote pairs of nodes within the network \cite{schoute.etal_2016, acin.etal_2007, perseguers.etal_2008, pirandola_2019a, shi.qian_2020, meignant.etal_2019}. 

Complex network topology is an emergent property observed in diverse real world networks. Complex networks prominently exhibit a heavy-tailed degree distribution, approximated by a power-law as
\begin{equation}
    P(k) \propto k^{-\gamma}
\end{equation}
with $2<\gamma<3$, where $k$ is the degree of the node and $P(k)$ is the fraction of nodes having degree $k$ \cite{barabasi_2016}. Another key feature of network complexity is the small-world property, stating that other nodes can be reached in a few steps from any given node: the diameter $d$ of the network (the maximum shortest path length between any two nodes) increases slower than a power-law, for example $d \propto \ln{N}$ for a network of $N$ nodes. Complex topology has been shown to be essential to the proper functioning of the classical internet, for instance by making it robust to the random failure of computers or routers \cite{albert.etal_2000a}.

Recent findings highlight the need to investigate the possibility of complex topology in a quantum internet. For example, complex quantum communication networks exhibit promising robustness to random failures of noisy quantum-repeater nodes \cite{coutinho.etal_2022} (though in the different framework of quantum spin models on imprinted networks, complexity does not necessarily provide robustness \cite{sundar.etal_2021}). Additionally, satellite-based quantum networks naturally have the small world property, meaning only a few entanglement swaps are needed to enable communication between any pair of nodes \cite{brito.etal_2021}. This is advantageous due to reduced usage of quantum resources. Another advantage to complex topology is that existing optical cables of the classical internet could be utilized for quantum communication, likely as part of a hybrid architecture along with more expensive cryo-cables, thus providing incentive to eventually match (at least partially) the complex topology of the classical internet \cite{rabbie.etal_2022, vanleent.etal_2022, luo.etal_2022, cacciapuoti.etal_2020}. 

However, there are also anticipated difficulties with creating quantum communication networks that mimic the complexity of the classical internet. For one, unlike in the case of satellite-based quantum networks, quantum networks based on optical fibers do not necessarily have the small world property due to photonic losses limiting transmission length \cite{brito.etal_2020}. There are also concerns that low network densities of complex networks may limit end-to-end transmission capacity of quantum information, and that there is vulnerability to targeted attacks, just as in the classical internet \cite{zhang.zhuang_2021, zhuang.zhang_2021}. By default, existing proposals for the creation of quantum communication networks avoid these issues by imposing a simple, grid-like network structure \cite{kimble_2008, kozlowski.wehner_2019}. Further, existing quantum routing protocols often assume simple grid-like or ring-like network structure \cite{schoute.etal_2016, acin.etal_2007, perseguers.etal_2008, pirandola_2019a}, with Ref.~\cite{shi.qian_2020} as a recent exception. As such, there is a need for model systems in which the impact of complex topology on quantum communication can be studied.

In this paper, we present a proof-of-concept model to create quantum networks of substantial complexity in the context of spins on a lattice. As building blocks, we consider entangled spin clusters of interacting quantum spin models on a two-dimensional lattice, where the ground state factorizes into independent spin clusters (corresponding to magnetic domains). Our work is in part inspired by Ref.~\cite{bianconi_2013}, where the author studied superconductivity by viewing clusters as nodes and adding a link between clusters if they were sufficiently close together. In contrast, here we choose regions of the lattice as nodes, and link two nodes if a spin cluster overlaps both node regions and no other node regions, according to the process described in Section \ref{sec:RTIM}. As a prominent example, we focus on ground state GHZ clusters of the two-dimensional random transverse-field Ising model (RTIM), but our results apply more generally, as discussed later. In Section \ref{sec:complexity} we analyze the resulting quantum networks and show that they have a similar complexity to that observed for the classical internet. We discuss the results and conclude by outlining further research directions in Section \ref{sec:discussion}.

%%%%%%%%%%%%%%%%
% QUANTUM NETWORK CONSTRUCTION
%%%%%%%%%%%%%%%%

\section{Quantum network construction} 
\label{sec:RTIM}

In this work, ground-state spin clusters of the RTIM serve as links for constructing complex quantum networks. As an overview of our construction, network nodes are chosen to be connected regions of lattice sites, with a broad distribution of sizes, for example as shown in Figure \ref{fig:panel_1}c. Then, nodes are connected by a quantum link if they both have a site belonging to the same spin cluster, and that cluster has no sites in another node, as shown in Figure \ref{fig:panel_1}b.

The two-dimensional RTIM is specified by the Hamiltonian
\begin{equation}
    \label{RTIM}
    \mathcal{H} = - \sum_{\langle ij \rangle} J_{ij} \sigma_i^x \sigma_j^x - \sum_i h_i \sigma_i^z\;,
\end{equation}
where each $\sigma_i^{\alpha}$ is the spin-$1/2$ Pauli matrix in the $\alpha$ direction for a spin at site $i$ of a $L \times L$ square lattice with periodic boundary conditions. The label $\langle i j \rangle$ indicates that the sum is taken over neighboring sites $i$ and $j$ on the lattice, which are coupled by bonds of random strength $J_{ij}\geq 0$, each drawn independently from the uniform distribution on the unit interval. Each site $i$ has a transverse external magnetic field of strength $h_i$. The relevant physics of the RTIM remains unchanged for any non-singular distribution of $h_i$ and $J_{ij}$ as long as at least the $h_i$ fields or the $J_{ij}$ bonds are chosen randomly \cite{igloi.monthus_2005, igloi.monthus_2018}. Therefore, to simplify the model we consider a `fixed-$h$' model with a uniform magnetic field $h_i=h$ for all sites $i$ \cite{kovacs.igloi_2011, kovacs.igloi_2010a}, unless otherwise stated. 

The RTIM is a paradigmatic example of a system which can undergo a quantum phase transition at zero temperature as the quantum control parameter $\theta = \ln{h}$ is tuned past its critical value $\theta_c$ \cite{sachdev_2000, sachdev_2011}. Below $\theta_c$ the ground state has a macroscopic spin cluster ordered by the couplings $J_{ij}$, and above $\theta_c$ the ground state is given by small spin clusters aligned independently. In contrast, at the critical point $\theta_c = -0.17034(2)$, the spin clusters have a broad size-distribution and are self-similar fractal-like objects \cite{kovacs.igloi_2011, kovacs.igloi_2010a}. Many properties of the critical RTIM are universal \cite{igloi.monthus_2005, igloi.monthus_2018}, that is, they are independent of the form of the disorder in $h$ and $J$, as well as of the type of the 2D lattice. 

The ground state of the RTIM can be conveniently determined using the strong disorder renormalization group (SDRG) method \cite{ma.etal_1979, dasgupta.ma_1980, igloi.monthus_2005, igloi.monthus_2018}, which is asymptotically exact in the vicinity of the critical point \cite{danielreview} as demonstrated in both two and higher dimensions \cite{motrunich.etal_2000, lin.etal_2000, karevski.etal_2001, lin.etal_2007, yu.etal_2008, 2dRG, kovacs.igloi_2010a, ddRG, kovacs.igloi_2011, pich.etal_1998}. We used an efficient SDRG algorithm, which runs in $O(N\log N)$ time for $N=L^2$ sites, to generate instances of RTIM ground state clusters at the critical point \cite{ddRG,kovacs.igloi_2011}. During the SDRG method, the largest local terms in the Hamiltonian are successively eliminated and new Hamiltonians are generated through a perturbation calculation \cite{kovacs.igloi_2011}. After decimating all degrees of freedom, the ground state of the RTIM is found to be a collection of independent ferromagnetic clusters of various size, each cluster being in a GHZ state
\begin{equation}
    \frac{1}{\sqrt{2}} \big( \underbrace{\ket{\uparrow \cdots \uparrow}}_{n \ times} + \underbrace{\ket{\downarrow \cdots \downarrow}}_{n \ times} \big)
\end{equation}
where $n$ is the number of spins in the cluster. In practice, we considered lattices up to size $L = 4096$, with at least $16$ instances at each size. An advantage of  using the critical point is that it leads to large network size after the network construction is carried out (see Figure \ref{fig:panel_4}d), while the SDRG remains asymptotically exact. 
Note that the ground state of the RTIM factorizes into a collection of independent GHZ clusters even outside the critical point \cite{kovacs.juhasz_2020}, indicating that off-critical RTIMs could also be used.

Let us first consider one of the network nodes, a spatially localized region $\mathcal{A}$ of the RTIM lattice. The entanglement entropy
\begin{equation} \label{entanglement entropy}
    S = - \mathrm{Tr}(\rho_{\mathcal{A}} \log_2 \rho_{\mathcal{A}}), 
\end{equation}
which is the von Neumann entropy of the reduced density matrix $\rho_{\mathcal{A}}$, provides a quantification of the entanglement between $\mathcal{A}$ and the rest of the lattice $\mathcal{A}^c$. In the RTIM, the entanglement entropy is simply the number of clusters with spins in both $\mathcal{A}$ and $\mathcal{A}^c$ \cite{refael_moore04, lin.etal_2007, yu.etal_2008, EPL}. As a special case of the `area law', the entanglement entropy is on average proportional to the surface area (boundary length) of $\mathcal{A}$ \cite{area}. As a consequence, regions with a larger boundary have a larger capacity to establish connections, leading to a proportionally larger expected degree. Therefore, to achieve a heavy-tailed degree distribution for the eventual quantum networks, we aimed to choose node regions with a broad distribution of surface area. As a simple choice, we selected discretized circles of varying sizes in the RTIM lattice as our regions, as shown in Figure \ref{fig:panel_1}c. We chose to sample the circles' radii from a power-law distribution with exponent $\gamma_{\mathrm{radius}} = 2.67$. (Other values of $\gamma_{\mathrm{radius}}$ between $2$ and $3$ give qualitatively similar results.) A minimum radius of $2$ lattice units was used to ensure the disks would consist of at least a few lattice sites. To maximize connectivity, we packed the circles relatively densely on the lattice using the method in Ref.~\cite{wang.etal_2006}. This method sequentially places circles in an outward-spiralling manner so that a new circle is tangent to one or more of the already placed circles. Note that this method does not attempt to solve the challenging problem of maximizing the packing density, nor is it a uniformly random placement of the circles. In practice, disks were added until a fixed proportion, here chosen to be $0.3$, of the lattice sites were covered, as illustrated in Figure \ref{fig:panel_1}c.

Now consider two disjoint regions $\mathcal{A}$ and $\mathcal{B}$ serving as network nodes, as well as the rest of the lattice, as shown in Figure \ref{fig:panel_1}b. As we partition the system into more than two subsystems, it becomes challenging to quantify quantum entanglement between pairs of subsystems \cite{szalay}, as the entanglement entropy is no longer applicable. As an alternative, one can consider the logarithmic negativity, which is an upper bound for the distillable entanglement, although it is notoriously difficult to compute in practice \cite{EN5,EN6}. In the RTIM, the entanglement negativity is non-zero if and only if $\mathcal{A}$ and $\mathcal{B}$ share a spin cluster that has no sites outside of these two regions \cite{calabrese, zou.etal_2022}. As the most strict construction, we could consider only such clusters as communication channels between the nodes. However, this is overly restrictive in the context of quantum networks, as any cluster that has sites in both regions can be used for communication, as long as the other sites in the cluster remain unchanged. As it is reasonable to assume that spins are only manipulated inside the nodes, we chose to add a quantum link between $\mathcal{A}$ and $\mathcal{B}$ if they both have a site belonging to the same spin cluster, and that cluster has no sites in any other node. Note that in our construction $\mathcal{A}$ and $\mathcal{B}$ are typically adjacent and at least some of the shared clusters are relatively small, meaning that there is a negligible difference between these two alternative cluster selection protocols.

If the resulting network is disconnected (as in Figure \ref{fig:panel_1}d), we considered only the largest connected component (LCC) of the network. In other words, each connected component is a separate quantum communication network, and we only consider the largest here.

%%%%%%%%%%%%%%%%
% TOPOLOGICAL ANALYSIS OF QUANTUM NETWORKS
%%%%%%%%%%%%%%%%

\section{Topological analysis of quantum networks}
\label{sec:complexity}

\begin{figure}[t]
    \begin{center}
        \includegraphics{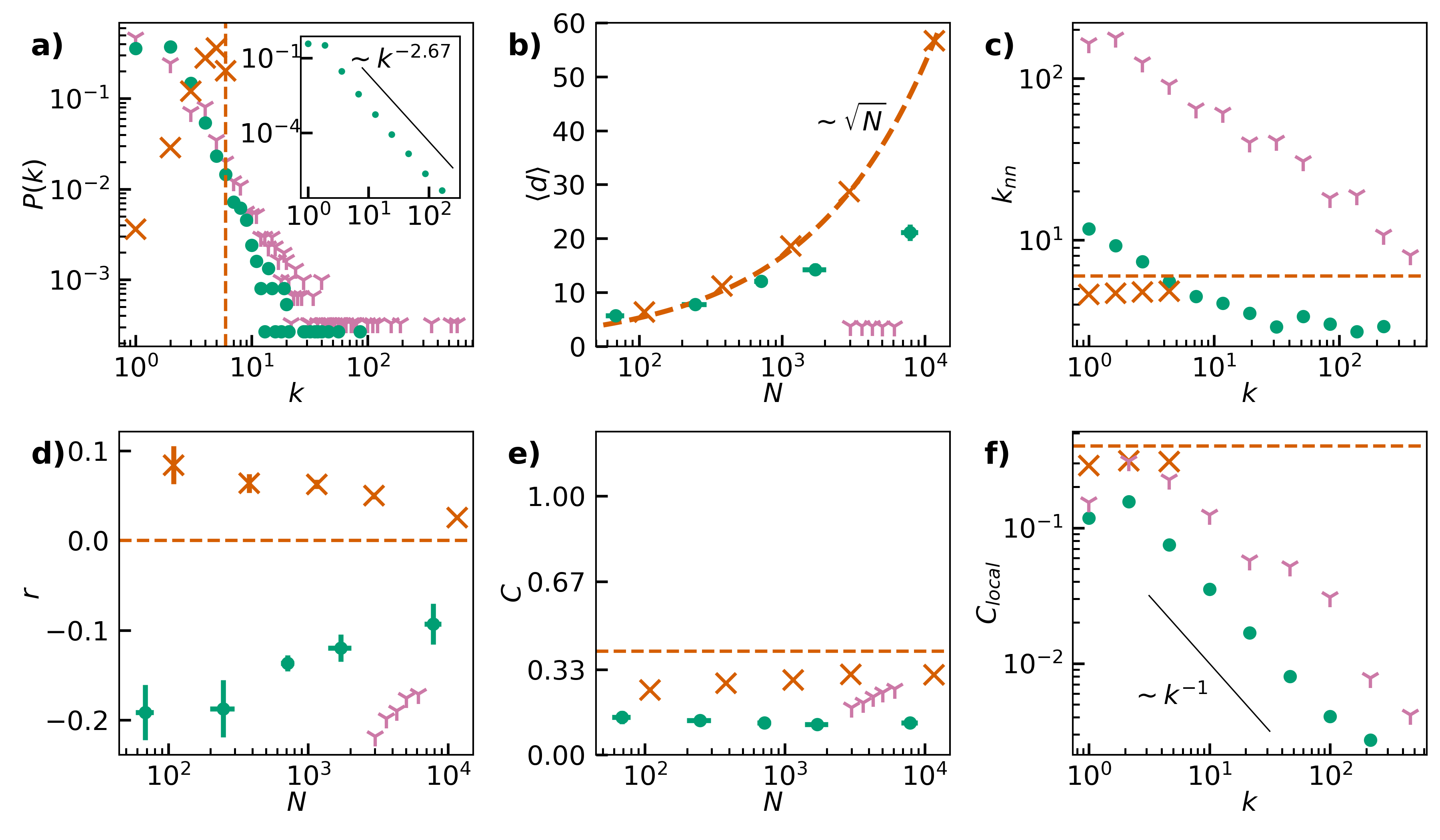}
	    \caption{\textbf{Analysis of quantum network topology.}
	             \textbf{a)} Degree distribution of a quantum network created with heterogeneously sized (green) or uniformly sized (dark orange) circular nodes on a large RTIM lattice, and the degree distribution of the classical internet at the autonomous systems level on 11/08/1997 (purple) \cite{meyer_}. The dashed line is the expectation for a triangular grid. Inset: The average degree distribution of $16$ quantum networks appears to obey a power-law with exponent approximately the same as that of the node size distribution.
	             \textbf{b)} Average shortest path length of networks from a few RTIM lattice sizes. Error bars represent standard error of the mean.
	             \textbf{c)} Average nearest neighbor degree as a function of node degree.
	             \textbf{d)} Degree correlation coefficients at a few sizes.
	             \textbf{e)} Global clustering coefficient.
	             \textbf{f)} Local clustering coefficient vs.~degree.
	            }
	    \label{fig:panel_3}
    \end{center}
\end{figure}

Visualizations of quantum networks constructed using the method of the previous section (Figure \ref{fig:panel_1}e) indicate nontrivial topological features, including highly connected hubs and clustering. This suggests the presence of network complexity, which we verify here numerically. For comparison, we also consider grid-like benchmark networks lacking complexity: quantum networks constructed using the method of Section \ref{sec:RTIM} but with uniformly sized nodes. In this grid-like benchmark, the radius of the nodes was set so that we have the same overall coverage of the RTIM lattice as in our complex quantum networks. The nodes were arranged in a hexagonal packing on the lattice, and as such the resulting networks are expected to approximate a triangular grid. We also compare to the topology of the classical internet at the level of autonomous systems from $1997$ until $2000$ \cite{meyer_}. From $733$ available snapshots we selected five representative networks to serve as a basis for comparison, the earliest of which is visualized in Figure \ref{fig:panel_1}f.

We first check that the quantum networks have a heavy-tailed degree distribution, a hallmark property of network complexity. As the node size distribution obeys a power-law, the expectation based on the area law is that the degree distribution obeys a power-law with the same exponent, under the assumption that node positions were chosen truly randomly. Indeed, Figure \ref{fig:panel_3}a shows a degree distribution of a typical quantum network constructed using the methods of Section \ref{sec:RTIM}, with a power-law exponent approximately the same as the one by which the node sizes were chosen. A slight difference in the power-law exponent may be attributed to the non-uniformly random selection of the node regions. The degree distribution is qualitatively similar to that of a representative network of the classical internet at the autonomous systems level, while contrasting with grid-like benchmark networks. The inset in Figure \ref{fig:panel_3}a displays an even more clear power-law behavior, upon averaging the degree distribution for a set of quantum networks.
% We could mention here a contrast to satellite-based quantum comunication networks with a log-normal degree distribution.

Another key feature of complex networks is the small-world property. Here, we used the average shortest path length $\langle d \rangle$ between two nodes as a proxy for the diameter of the quantum networks, shown in Figure \ref{fig:panel_3}b for quantum networks from different RTIM system sizes. Though the average path length of the quantum networks is larger than that of the classical internet networks, it is evident that it scales much more slowly than in the grid-like network, which has $\langle d \rangle  \propto \sqrt{N}$. In fact, quantum networks are consistent with a logarithmic relationship $\langle d \rangle \propto \ln{N}$, indicating that these networks have the small-world property, although with a somewhat larger diameter than the classical internet.

A closer examination of individual quantum networks reveals that spatially large nodes do not always have a high degree. Depending on the number and size of the surrounding nodes, a large node can often end up with only a few connections, especially if the node is on the periphery of the node packing configuration, or if it is next to other large nodes. We therefore checked for the presence of degree correlations between linked nodes. In Figure \ref{fig:panel_3}c the average degree of a node's nearest neighbors is shown as a function of the node's own degree. Just like in the classical internet, the negative slope indicates that the quantum networks are disassortative, meaning high degree nodes tend to connect to low degree nodes and vice versa. This contrasts with the grid-like benchmark networks, which exhibit approximately neutral behavior. The observed network disassortativity can be quantified by the degree correlation coefficient $r$, which is plotted against the network size in Figure \ref{fig:panel_3}d. A correlation coefficient $r=0$ means no degree correlations, while negative values indicate disassortativity. Both the quantum networks and the classical internet exhibit disassortativity for all $N$, but they tend to become less disassortative as $N$ increases. In contrast, grid-like networks have asymptotically no degree correlations. For small sizes, low degree nodes often appear next to each other in areas dominated by a large cluster in the RTIM, as well as on the periphery of the grid, leading to slightly positive $r$ values. 

Degree correlations capture patterns at the level of pairs of connected nodes, but it is often useful to go one step beyond and check patterns of three nodes. The simplest such measurement is the global clustering coefficient, given by
\begin{equation}
    C \equiv \frac{3 \cdot \text{Number of complete triangles}}{\text{Number of connected triplets of nodes}} 
    = \frac{\sum_{i,j,k} A_{ij} A_{jk} A_{ki}}{\sum_i k_i (k_i - 1)}
\end{equation}
where $A$ is the adjacency matrix of the network. The global clustering coefficient of quantum networks of various sizes are shown in Figure \ref{fig:panel_3}e. A perfect triangular lattice would achieve $C=0.4$, somewhat above the grid-like benchmark networks that often miss some connections. The clustering coefficient of classical internet networks falls into the range spanned by the heterogeneous and grid-like benchmark networks.

Beyond the global clustering coefficient, we can determine if the network exhibits hierarchical modularity, in which low degree nodes tend to exist in dense communities while high degree nodes connect disparate communities, by examining the relationship between the local clustering coefficient and node degree. The local clustering coefficient of node $i$ is 
\begin{equation}
        C_{\mathrm{local},i} = \frac{\sum_{j,k} A_{ij} A_{jk} A_{ki}}{ k_i (k_i - 1)}.
\end{equation}
Hierarchical network structure is indicated by a local clustering coefficient that decays as the node degree increases. As shown in Figure \ref{fig:panel_3}f, we indeed observe that $C_{\mathrm{local}}$ decays with $k$ in both the quantum networks and the classical internet network. Further, the local clustering coefficient of the quantum networks obeys the relation $C_{\mathrm{local}} \sim k^{-1}$ well, indicating agreement with the hierarchical network model \cite{ravasz.barabasi_2003}. Thus the quantum networks share the hierarchical nature of the classical internet, in stark contrast to grid-like networks with no hubs.

%%%%%%%%%%%%%%%%
% DISCUSSION AND CONCLUSIONS
%%%%%%%%%%%%%%%%

\section{Discussion and conclusions}
\label{sec:discussion}

In this work, we have introduced a theoretical method to create model quantum communication networks on a lattice of spins using entangled clusters as quantum communication channels. Specifically, we have shown that ground state GHZ clusters of the fixed-$h$ critical RTIM can be used as quantum communication links between local regions of spins, yielding quantum networks with substantial network complexity. As a key implication of the complex topology, the resulting networks are expected to be robust against random node and link failures, while being vulnerable against targeted attacks of the highly connected hubs, similarly to the classical internet and satellite-based quantum communication networks \cite{brito.etal_2021}.

\begin{figure}[t]
    \begin{center}
        \includegraphics{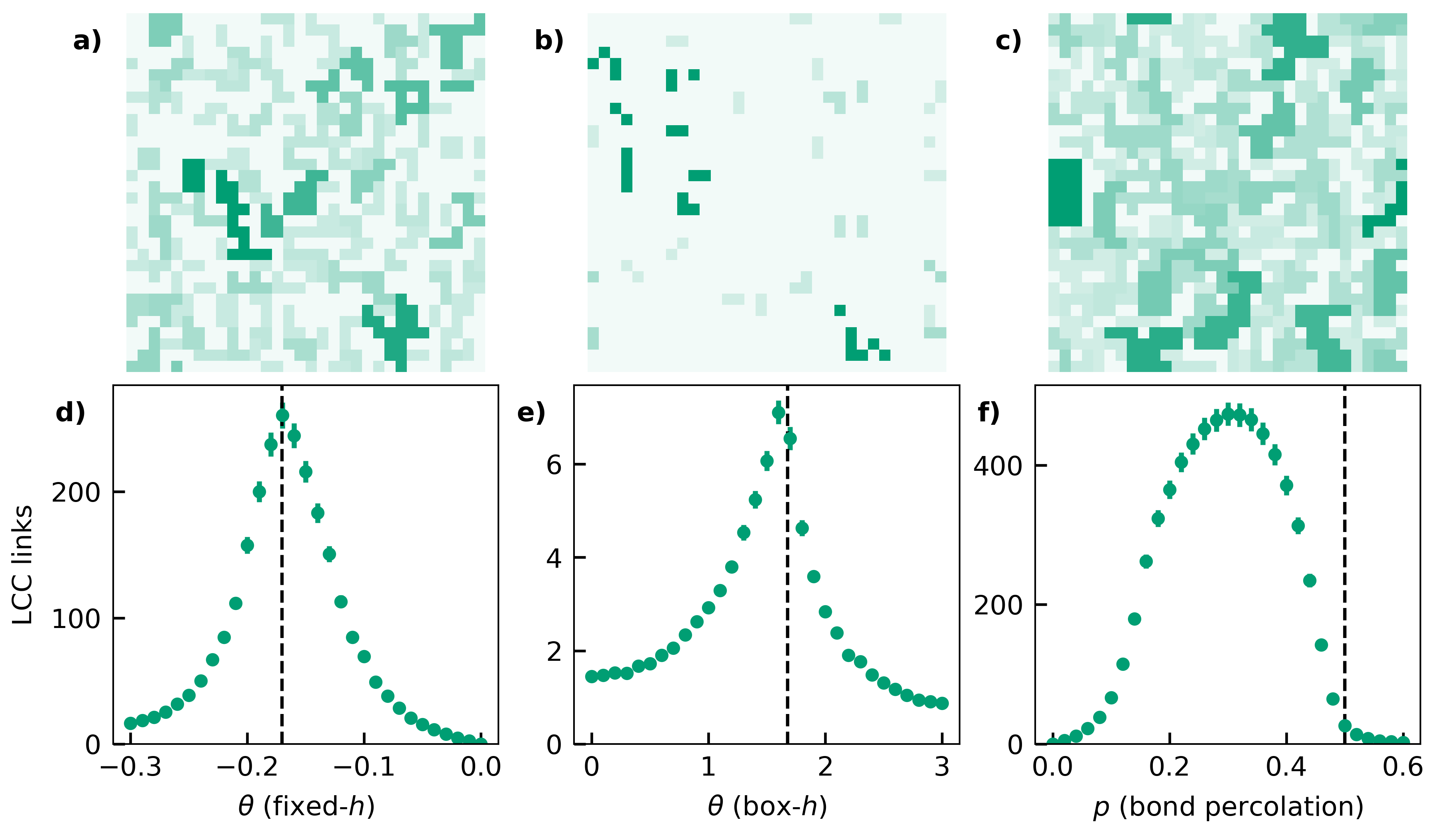}
	    \caption{
	        \textbf{Spin clusters in variants of the RTIM used to create quantum networks.}
	        \textbf{a)} Fixed-$h$ RTIM ground state clusters at the critical $\theta$. Clusters are colored by size, and may be disconnected \cite{EPL, kovacs.juhasz_2020}.
	        \textbf{b)} Box-$h$ RTIM ($h_i$ uniformly distributed in $(0, h]$) ground state clusters at critical $\theta$.
	        \textbf{c)} Bond percolation clusters, which are ground state clusters of the diluted RTIM ($J_{ij}=J\gg h$ with probability $p$, $J_{ij}=0$ otherwise) \cite{senthil.sachdev_1996, harris_1974, stinchcombe_1981, dossantos_1982}. Here $p=0.3$.
	        \textbf{d-f)} For each system (here on a $512 \times 512$ square lattice), there is an optimal value of a cluster control parameter which produces the largest LCC (averaged over 256 samples; error bars represent standard error of the mean). In the fixed-$h$ RTIM the optimal value of $\theta$ approximately coincides with the critical point (dashed lines), but this does not hold in general.
	            }
	    \label{fig:panel_4}
    \end{center}
\end{figure}

Our approach can be further generalized, starting with using a non-critical RTIM. In Figure \ref{fig:panel_4}a,d we have performed the quantum network construction of Section \ref{sec:RTIM} with the control parameter $\theta$ at various values away from $\theta_c$, and plotted the average size (number of links) of the LCC of the resulting network. For $\theta$ near but not at $\theta_c$, the network construction still produces quantum networks of substantial size, with similar properties and complexity to that found in Section \ref{sec:complexity}. The network size achieves a maximum for $\theta = \theta_c$, although this is a coincidence as criticality is not required in our network construction. What we need is a large number of local clusters, which happens to occur in the vicinity of the critical point for the fixed-$h$ disorder distribution. To show this, in Figure \ref{fig:panel_4}b,e we used the box-$h$ RTIM, where the $h_i$ are uniformly distributed in $(0, h]$, with the quantum network construction of Section \ref{sec:RTIM}. In this case we observe a slight discrepancy between the critical point and the optimal $\theta$ parameter value. Note that in this system the smaller size of clusters led to much smaller networks on average, but we still expect to see complexity at larger scales.

We may consider further variants of the RTIM, such as the diluted RTIM ($J_{ij}=J\gg h$ with probability $p$, $J_{ij}=0$ otherwise) \cite{senthil.sachdev_1996, harris_1974, stinchcombe_1981, dossantos_1982}. This system has connected ground state clusters in the shape of bond percolation clusters, shown in Figure \ref{fig:panel_4}d,f, and belongs to a different universality class than the fixed-$h$ and box-$h$ RTIM \cite{EPL, kovacs.juhasz_2020}. Yet, our network construction still leads to large networks of substantial complexity. More generally, the quantum network construction can be extended to any system with a ground state that factorizes into independent spin clusters. Such systems include the random Potts and clock models \cite{senthil.majumdar_1996} as well as the Ashkin-Teller model \cite{carlon.etal_2001}.

As a further extension, spin clusters of different layers could serve both as nodes and links. For example, one layer of the diluted critical RTIM (with percolation clusters) could define the nodes, while another layer could provide the links. As clusters in the critical diluted RTIM have a broad size distribution, much like the circular nodes chosen earlier, they would naturally lead to a broad degree distribution. The volume distribution exponent is given by $\tau = 1 + D / D_f$, where $D=2$ is the dimension of the lattice and $D_f = 91 / 48$ is the fractal dimension of percolation clusters \cite{stauffer.aharony_2017}.

Our network construction can lead to even more dense and potentially more complex networks by using multiple layers of spin clusters for links. This idea leads to multilayer networks \cite{kivela.etal_2014}, where the union of the links coming from different layers can increase the topological complexity further, no longer resulting in the current, quasi-planar graphs. In both a single-layer and multi-layer construction, we can construct a weighted network, where the link weight is the multiplicity of the established quantum channels between each node pair. While for a purely topological analysis such weights are irrelevant, they can be of key importance for practical quantum communication protocols \cite{coutinho.etal_2022}.

Note that scale-free spatial networks would traditionally require long range connections \cite{coutinho.etal_2022}. As our design includes spatially extended nodes, scale-free quantum networks can be achieved even with only short range connections, like in the optimal off-critical construction in Figure \ref{fig:panel_4}f. Our quantum network construction could also generalize to higher dimensions. In 3D, the area law means that the degree distribution is coupled to the distribution of the area of the subsystems instead of their linear extent. Hence the degree distribution is expected to obey a power law with an exponent one larger than the exponent of the linear size distribution. Unlike the networks from the 2D construction, networks from the 3D construction would not be nearly planar, and would likely have higher clustering coefficient. In contrast, the same construction would not work in the 1D RTIM: Since the boundary of a connected region is constant, all connected node regions are expected to have equal average degree. A caveat is that for a 1D RTIM at criticality, there are logarithmic corrections to the area law \cite{holzhey, vidal, Calabrese_Cardy04}; however, it would be unwieldy to create power law distributions from merely logarithmic corrections.

In conclusion, we have presented a way to create model complex quantum networks on a lattice of spins. As the first application, our work centers around the critical two-dimensional RTIM, but generalizations to other interacting spin systems that admit entangled ground state clusters are apparent. Our model serves as an accessible generative framework for further investigations on network complexity in the emerging quantum internet. Further, our results could motivate experimental work to create a spatially small but many-node complex quantum communication network using a magnetic solid that admits magnetic domains in the form of spin clusters. We believe that in addition to ongoing and future experiments with a few nodes separated at large distances  \cite{peev.etal_2009, sasaki.etal_2011, liao.etal_2018, wehner.etal_2018}, we also need small-scale experiments with many nodes to explore the implications of complex network topologies. This is an exciting time to explore the question of quantum network topology, influencing how the quantum internet will be shaped. As of now, it is unclear if the emerging quantum internet will acquire a complex network structure. Even if the quantum internet falls into a novel class of complex networks substantially different from those observed in classical systems, models like the one presented here may be valuable to understand how network properties arise.

%%%%%%%%%%%%%%%%
% ACKNOWLEDGEMENTS, ETC
%%%%%%%%%%%%%%%%

\section*{Acknowledgements}
This work has been supported by the Baker Faculty Grant of the Weinberg College of Arts and Sciences, Northwestern University, 2020. We acknowledge support from the JTF project \emph{The Nature of Quantum Networks} (ID 60478). In addition, R.T.C.~Chepuri was supported by the Northwestern University SURG-Advanced in 2021. We are thankful for Ginestra Bianconi for useful discussions. We also thank Bingjie Hao, Anastasiya Salova, and Helen S.~Ansell for insightful feedback on the manuscript.

\section*{Author contributions}
I.A.K.~developed the initial concept and supervised the research. R.T.C.C.~performed the network simulations and numerical analyses. All authors contributed to the design and writing of the manuscript.

\section*{Competing interests}
The authors declare no competing interests.

\section*{Data availability}
The data that support the findings of this study are available upon reasonable request.

\section*{Additional information}
Correspondence and requests for materials should be addressed to I.A.~Kov\'acs.

%%%%%%%%%%%%%%%%
% BIBLIOGRAPHY
%%%%%%%%%%%%%%%%

\printbibliography

@article{2dRG,
  title = {Critical Behavior and Entanglement of the Random Transverse-Field {{Ising}} Model between One and Two Dimensions},
  author = {Kov{\'a}cs, Istv{\'a}n A. and Igl{\'o}i, Ferenc},
  year = {2009},
  month = dec,
  journal = {Phys. Rev. B},
  volume = {80},
  number = {21},
  pages = {214416},
  publisher = {{American Physical Society}},
  doi = {10.1103/PhysRevB.80.214416}
}

@article{acin.etal_2007,
  title = {Entanglement Percolation in Quantum Networks},
  author = {Ac{\'i}n, Antonio and Cirac, J. Ignacio and Lewenstein, Maciej},
  year = {2007},
  month = apr,
  journal = {Nature Phys},
  volume = {3},
  number = {4},
  pages = {256--259},
  publisher = {{Nature Publishing Group}},
  issn = {1745-2481},
  doi = {10.1038/nphys549},
  copyright = {2007 Nature Publishing Group},
  langid = {english}
}

@article{albert.etal_2000a,
  title = {Error and Attack Tolerance of Complex Networks},
  author = {Albert, R{\'e}ka and Jeong, Hawoong and Barab{\'a}si, Albert-L{\'a}szl{\'o}},
  year = {2000},
  month = jul,
  journal = {Nature},
  volume = {406},
  number = {6794},
  pages = {378--382},
  publisher = {{Nature Publishing Group}},
  issn = {1476-4687},
  doi = {10.1038/35019019},
  copyright = {2000 Macmillan Magazines Ltd.},
  langid = {english}
}

@article{area,
  title = {Colloquium: {{Area}} Laws for the Entanglement Entropy},
  shorttitle = {Colloquium},
  author = {Eisert, J. and Cramer, M. and Plenio, M. B.},
  year = {2010},
  month = feb,
  journal = {Rev. Mod. Phys.},
  volume = {82},
  number = {1},
  pages = {277--306},
  publisher = {{American Physical Society}},
  doi = {10.1103/RevModPhys.82.277}
}

@book{barabasi_2016,
  title = {Network {{Science}}},
  author = {Barab{\'a}si, Albert-L{\'a}szl{\'o}},
  year = {2016},
  month = aug,
  edition = {1st edition},
  publisher = {{Cambridge University Press}},
  address = {{Cambridge, United Kingdom}},
  isbn = {978-1-107-07626-6},
  langid = {english}
}

@article{biamonte.etal_2019,
  title = {Complex Networks from Classical to Quantum},
  author = {Biamonte, Jacob and Faccin, Mauro and De Domenico, Manlio},
  year = {2019},
  month = may,
  journal = {Commun Phys},
  volume = {2},
  number = {1},
  pages = {1--10},
  publisher = {{Nature Publishing Group}},
  issn = {2399-3650},
  doi = {10.1038/s42005-019-0152-6},
  copyright = {2019 The Author(s)},
  langid = {english}
}

@article{bianconi_2013,
  title = {Superconductor-Insulator Transition in a Network of 2d Percolation Clusters},
  author = {Bianconi, Ginestra},
  year = {2013},
  month = jan,
  journal = {EPL},
  volume = {101},
  number = {2},
  pages = {26003},
  issn = {0295-5075},
  doi = {10.1209/0295-5075/101/26003},
  langid = {english}
}

@article{brito.etal_2020,
  title = {Statistical {{Properties}} of the {{Quantum Internet}}},
  author = {Brito, Samura{\'i} and Canabarro, Askery and Chaves, Rafael and Cavalcanti, Daniel},
  year = {2020},
  month = may,
  journal = {Phys. Rev. Lett.},
  volume = {124},
  number = {21},
  pages = {210501},
  publisher = {{American Physical Society}},
  doi = {10.1103/PhysRevLett.124.210501}
}

@article{brito.etal_2021,
  title = {Satellite-{{Based Photonic Quantum Networks Are Small-World}}},
  author = {Brito, Samura{\'i} and Canabarro, Askery and Cavalcanti, Daniel and Chaves, Rafael},
  year = {2021},
  month = jan,
  journal = {PRX Quantum},
  volume = {2},
  number = {1},
  pages = {010304},
  publisher = {{American Physical Society}},
  doi = {10.1103/PRXQuantum.2.010304}
}

@article{cacciapuoti.etal_2020,
  title = {Quantum {{Internet}}: {{Networking Challenges}} in {{Distributed Quantum Computing}}},
  shorttitle = {Quantum {{Internet}}},
  author = {Cacciapuoti, Angela Sara and Caleffi, Marcello and Tafuri, Francesco and Cataliotti, Francesco Saverio and Gherardini, Stefano and Bianchi, Giuseppe},
  year = {2020},
  month = jan,
  journal = {IEEE Netw.},
  volume = {34},
  number = {1},
  pages = {137--143},
  issn = {1558-156X},
  doi = {10.1109/MNET.001.1900092}
}

@article{calabrese,
  title = {Entanglement Negativity in Random Spin Chains},
  author = {Ruggiero, Paola and Alba, Vincenzo and Calabrese, Pasquale},
  year = {2016},
  month = jul,
  journal = {Phys. Rev. B},
  volume = {94},
  number = {3},
  pages = {035152},
  publisher = {{American Physical Society}},
  doi = {10.1103/PhysRevB.94.035152}
}

@article{Calabrese_Cardy04,
  title = {Entanglement Entropy and Quantum Field Theory},
  author = {Calabrese, Pasquale and Cardy, John},
  year = {2004},
  month = jun,
  journal = {J. Stat. Mech.},
  number = {06},
  pages = {P06002},
  publisher = {{IOP Publishing}},
  issn = {1742-5468},
  doi = {10.1088/1742-5468/2004/06/P06002},
  langid = {english}
}

@article{carlon.etal_2001,
  title = {Disorder {{Induced Cross-Over Effects}} at {{Quantum Critical Points}}},
  author = {Carlon, Enrico and Lajk{\'o}, P{\'e}ter and Igl{\'o}i, Ferenc},
  year = {2001},
  month = dec,
  journal = {Phys. Rev. Lett.},
  volume = {87},
  number = {27},
  pages = {277201},
  publisher = {{American Physical Society}},
  doi = {10.1103/PhysRevLett.87.277201}
}

@article{chen.etal_2020,
  title = {Sending-or-{{Not-Sending}} with {{Independent Lasers}}: {{Secure Twin-Field Quantum Key Distribution}} over 509 Km},
  shorttitle = {Sending-or-{{Not-Sending}} with {{Independent Lasers}}},
  author = {Chen, Jiu-Peng and Zhang, Chi and Liu, Yang and Jiang, Cong and Zhang, Weijun and Hu, Xiao-Long and Guan, Jian-Yu and Yu, Zong-Wen and Xu, Hai and Lin, Jin and Li, Ming-Jun and Chen, Hao and Li, Hao and You, Lixing and Wang, Zhen and Wang, Xiang-Bin and Zhang, Qiang and Pan, Jian-Wei},
  year = {2020},
  month = feb,
  journal = {Phys. Rev. Lett.},
  volume = {124},
  number = {7},
  pages = {070501},
  publisher = {{American Physical Society}},
  doi = {10.1103/PhysRevLett.124.070501}
}

@article{cirac.etal_1997,
  title = {Quantum {{State Transfer}} and {{Entanglement Distribution}} among {{Distant Nodes}} in a {{Quantum Network}}},
  author = {Cirac, J. I. and Zoller, P. and Kimble, H. J. and Mabuchi, H.},
  year = {1997},
  month = apr,
  journal = {Phys. Rev. Lett.},
  volume = {78},
  number = {16},
  pages = {3221--3224},
  publisher = {{American Physical Society}},
  doi = {10.1103/PhysRevLett.78.3221}
}

@article{coutinho.etal_2022,
  title = {Robustness of Noisy Quantum Networks},
  author = {Coutinho, Bruno Coelho and Munro, William John and Nemoto, Kae and Omar, Yasser},
  year = {2022},
  month = apr,
  journal = {Commun Phys},
  volume = {5},
  number = {1},
  pages = {1--9},
  publisher = {{Nature Publishing Group}},
  issn = {2399-3650},
  doi = {10.1038/s42005-022-00866-7},
  copyright = {2022 The Author(s)},
  langid = {english}
}

@article{danielreview,
  title = {Phase Transitions and Singularities in Random Quantum Systems},
  author = {Fisher, Daniel S.},
  year = {1999},
  month = feb,
  journal = {Physica A: Statistical Mechanics and its Applications},
  series = {Proceedings of the 20th {{IUPAP International Conference}} on {{Statistical Physics}}},
  volume = {263},
  number = {1},
  pages = {222--233},
  issn = {0378-4371},
  doi = {10.1016/S0378-4371(98)00498-1},
  langid = {english}
}

@article{dasgupta.ma_1980,
  title = {Low-Temperature Properties of the Random {{Heisenberg}} Antiferromagnetic Chain},
  author = {Dasgupta, Chandan and Ma, Shang-Keng},
  year = {1980},
  month = aug,
  journal = {Phys. Rev. B},
  volume = {22},
  number = {3},
  pages = {1305--1319},
  publisher = {{American Physical Society}},
  doi = {10.1103/PhysRevB.22.1305}
}

@article{ddRG,
  title = {Infinite-Disorder Scaling of Random Quantum Magnets in Three and Higher Dimensions},
  author = {Kov{\'a}cs, Istv{\'a}n A. and Igl{\'o}i, Ferenc},
  year = {2011},
  month = may,
  journal = {Phys. Rev. B},
  volume = {83},
  number = {17},
  pages = {174207},
  publisher = {{American Physical Society}},
  doi = {10.1103/PhysRevB.83.174207}
}

@article{dossantos_1982,
  title = {The Pure and Diluted Quantum Transverse {{Ising}} Model},
  author = {{dos Santos}, R. R.},
  year = {1982},
  month = may,
  journal = {J. Phys. C: Solid State Phys.},
  volume = {15},
  number = {14},
  pages = {3141--3161},
  publisher = {{IOP Publishing}},
  issn = {0022-3719},
  doi = {10.1088/0022-3719/15/14/020},
  langid = {english}
}

@article{EN5,
  title = {Computable Measure of Entanglement},
  author = {Vidal, G. and Werner, R. F.},
  year = {2002},
  month = feb,
  journal = {Phys. Rev. A},
  volume = {65},
  number = {3},
  pages = {032314},
  publisher = {{American Physical Society}},
  doi = {10.1103/PhysRevA.65.032314}
}

@article{EN6,
  title = {Logarithmic {{Negativity}}: {{A Full Entanglement Monotone That}} Is Not {{Convex}}},
  shorttitle = {Logarithmic {{Negativity}}},
  author = {Plenio, M. B.},
  year = {2005},
  month = aug,
  journal = {Phys. Rev. Lett.},
  volume = {95},
  number = {9},
  pages = {090503},
  publisher = {{American Physical Society}},
  doi = {10.1103/PhysRevLett.95.090503}
}

@article{EPL,
  title = {Universal Logarithmic Terms in the Entanglement Entropy of 2d, 3d and 4d Random Transverse-Field {{Ising}} Models},
  author = {Kov{\'a}cs, I. A. and Igl{\'o}i, F.},
  year = {2012},
  month = mar,
  journal = {EPL},
  volume = {97},
  number = {6},
  pages = {67009},
  publisher = {{IOP Publishing}},
  issn = {0295-5075},
  doi = {10.1209/0295-5075/97/67009},
  langid = {english}
}

@article{fang.etal_2020,
  title = {Implementation of Quantum Key Distribution Surpassing the Linear Rate-Transmittance Bound},
  author = {Fang, Xiao-Tian and Zeng, Pei and Liu, Hui and Zou, Mi and Wu, Weijie and Tang, Yan-Lin and Sheng, Ying-Jie and Xiang, Yao and Zhang, Weijun and Li, Hao and Wang, Zhen and You, Lixing and Li, Ming-Jun and Chen, Hao and Chen, Yu-Ao and Zhang, Qiang and Peng, Cheng-Zhi and Ma, Xiongfeng and Chen, Teng-Yun and Pan, Jian-Wei},
  year = {2020},
  month = jul,
  journal = {Nat. Photonics},
  volume = {14},
  number = {7},
  pages = {422--425},
  publisher = {{Nature Publishing Group}},
  issn = {1749-4893},
  doi = {10.1038/s41566-020-0599-8},
  copyright = {2020 The Author(s), under exclusive licence to Springer Nature Limited},
  langid = {english}
}

@article{harris_1974,
  title = {Effect of Random Defects on the Critical Behaviour of {{Ising}} Models},
  author = {Harris, A. B.},
  year = {1974},
  month = may,
  journal = {J. Phys. C: Solid State Phys.},
  volume = {7},
  number = {9},
  pages = {1671--1692},
  publisher = {{IOP Publishing}},
  issn = {0022-3719},
  doi = {10.1088/0022-3719/7/9/009},
  langid = {english}
}

@article{holzhey,
  title = {Geometric and Renormalized Entropy in Conformal Field Theory},
  author = {Holzhey, Christoph and Larsen, Finn and Wilczek, Frank},
  year = {1994},
  month = aug,
  journal = {Nuclear Physics B},
  volume = {424},
  number = {3},
  pages = {443--467},
  issn = {0550-3213},
  doi = {10.1016/0550-3213(94)90402-2},
  langid = {english}
}

@article{igloi.monthus_2005,
  title = {Strong Disorder {{RG}} Approach of Random Systems},
  author = {Igl{\'o}i, Ferenc and Monthus, C{\'e}cile},
  year = {2005},
  month = jun,
  journal = {Physics Reports},
  volume = {412},
  number = {5},
  pages = {277--431},
  issn = {0370-1573},
  doi = {10.1016/j.physrep.2005.02.006},
  langid = {english}
}

@article{igloi.monthus_2018,
  title = {Strong Disorder {{RG}} Approach \textendash{} a Short Review of Recent Developments},
  author = {Igl{\'o}i, Ferenc and Monthus, C{\'e}cile},
  year = {2018},
  month = nov,
  journal = {Eur. Phys. J. B},
  volume = {91},
  number = {11},
  pages = {290},
  issn = {1434-6036},
  doi = {10.1140/epjb/e2018-90434-8},
  langid = {english}
}

@article{karevski.etal_2001,
  title = {Random Quantum Magnets with Broad Disorder Distribution},
  author = {Karevski, D. and Lin, Y.-C. and Rieger, H. and Kawashima, N. and Igl{\'o}i, F.},
  year = {2001},
  month = mar,
  journal = {Eur. Phys. J. B},
  volume = {20},
  number = {2},
  pages = {267--276},
  publisher = {{EDP Sciences}},
  issn = {1434-6028, 1434-6036},
  doi = {10.1007/PL00011100},
  copyright = {\textcopyright{} EDP Sciences, Societ\`a Italiana di Fisica, Springer-Verlag, 2001},
  langid = {english}
}

@article{kimble_2008,
  title = {The Quantum Internet},
  author = {Kimble, H. J.},
  year = {2008},
  month = jun,
  journal = {Nature},
  volume = {453},
  number = {7198},
  pages = {1023--1030},
  publisher = {{Nature Publishing Group}},
  issn = {1476-4687},
  doi = {10.1038/nature07127},
  copyright = {2008 Nature Publishing Group},
  langid = {english}
}

@article{kivela.etal_2014,
  title = {Multilayer Networks},
  author = {Kivel{\"a}, Mikko and Arenas, Alex and Barthelemy, Marc and Gleeson, James P. and Moreno, Yamir and Porter, Mason A.},
  year = {2014},
  month = sep,
  journal = {Journal of Complex Networks},
  volume = {2},
  number = {3},
  pages = {203--271},
  issn = {2051-1310},
  doi = {10.1093/comnet/cnu016}
}

@article{kovacs.igloi_2010a,
  title = {Renormalization Group Study of the Two-Dimensional Random Transverse-Field {{Ising}} Model},
  author = {Kov{\'a}cs, Istv{\'a}n A. and Igl{\'o}i, Ferenc},
  year = {2010},
  month = aug,
  journal = {Phys. Rev. B},
  volume = {82},
  number = {5},
  pages = {054437},
  publisher = {{American Physical Society}},
  doi = {10.1103/PhysRevB.82.054437}
}

@article{kovacs.igloi_2011,
  title = {Renormalization Group Study of Random Quantum Magnets},
  author = {Kov{\'a}cs, Istv{\'a}n A. and Igl{\'o}i, Ferenc},
  year = {2011},
  month = sep,
  journal = {J. Phys.: Condens. Matter},
  volume = {23},
  number = {40},
  pages = {404204},
  publisher = {{IOP Publishing}},
  issn = {0953-8984},
  doi = {10.1088/0953-8984/23/40/404204},
  langid = {english}
}

@article{kovacs.juhasz_2020,
  title = {Emergence of Disconnected Clusters in Heterogeneous Complex Systems},
  author = {Kov{\'a}cs, Istv{\'a}n A. and Juh{\'a}sz, R{\'o}bert},
  year = {2020},
  month = dec,
  journal = {Sci Rep},
  volume = {10},
  number = {1},
  pages = {21874},
  publisher = {{Nature Publishing Group}},
  issn = {2045-2322},
  doi = {10.1038/s41598-020-78769-2},
  copyright = {2020 The Author(s)},
  langid = {english}
}

@inproceedings{kozlowski.wehner_2019,
  title = {Towards {{Large-Scale Quantum Networks}}},
  booktitle = {Proc. {{Sixth Annu}}. {{ACM Int}}. {{Conf}}. {{Nanoscale Comput}}. {{Commun}}.},
  author = {Kozlowski, Wojciech and Wehner, Stephanie},
  year = {2019},
  month = sep,
  series = {{{NANOCOM}} '19},
  pages = {1--7},
  publisher = {{Association for Computing Machinery}},
  address = {{New York, NY, USA}},
  doi = {10.1145/3345312.3345497},
  isbn = {978-1-4503-6897-1}
}

@article{liao.etal_2017,
  title = {Satellite-to-Ground Quantum Key Distribution},
  author = {Liao, Sheng-Kai and Cai, Wen-Qi and Liu, Wei-Yue and Zhang, Liang and Li, Yang and Ren, Ji-Gang and Yin, Juan and Shen, Qi and Cao, Yuan and Li, Zheng-Ping and Li, Feng-Zhi and Chen, Xia-Wei and Sun, Li-Hua and Jia, Jian-Jun and Wu, Jin-Cai and Jiang, Xiao-Jun and Wang, Jian-Feng and Huang, Yong-Mei and Wang, Qiang and Zhou, Yi-Lin and Deng, Lei and Xi, Tao and Ma, Lu and Hu, Tai and Zhang, Qiang and Chen, Yu-Ao and Liu, Nai-Le and Wang, Xiang-Bin and Zhu, Zhen-Cai and Lu, Chao-Yang and Shu, Rong and Peng, Cheng-Zhi and Wang, Jian-Yu and Pan, Jian-Wei},
  year = {2017},
  month = sep,
  journal = {Nature},
  volume = {549},
  number = {7670},
  pages = {43--47},
  issn = {1476-4687},
  doi = {10.1038/nature23655},
  langid = {english},
  pmid = {28825707}
}

@article{liao.etal_2018,
  title = {Satellite-{{Relayed Intercontinental Quantum Network}}},
  author = {Liao, Sheng-Kai and Cai, Wen-Qi and Handsteiner, Johannes and Liu, Bo and Yin, Juan and Zhang, Liang and Rauch, Dominik and Fink, Matthias and Ren, Ji-Gang and Liu, Wei-Yue and Li, Yang and Shen, Qi and Cao, Yuan and Li, Feng-Zhi and Wang, Jian-Feng and Huang, Yong-Mei and Deng, Lei and Xi, Tao and Ma, Lu and Hu, Tai and Li, Li and Liu, Nai-Le and Koidl, Franz and Wang, Peiyuan and Chen, Yu-Ao and Wang, Xiang-Bin and Steindorfer, Michael and Kirchner, Georg and Lu, Chao-Yang and Shu, Rong and Ursin, Rupert and Scheidl, Thomas and Peng, Cheng-Zhi and Wang, Jian-Yu and Zeilinger, Anton and Pan, Jian-Wei},
  year = {2018},
  month = jan,
  journal = {Phys. Rev. Lett.},
  volume = {120},
  number = {3},
  pages = {030501},
  publisher = {{American Physical Society}},
  doi = {10.1103/PhysRevLett.120.030501}
}

@article{lin.etal_2000,
  title = {Numerical {{Renormalization Group Study}} of {{Random Transverse Ising Models}} in {{One}} and {{Two Space Dimensions}}},
  author = {Lin, Yu-Cheng and Kawashima, Naoki and Igl{\'o}i, Ferenc and Rieger, Heiko},
  year = {2000},
  month = apr,
  journal = {Progress of Theoretical Physics Supplement},
  volume = {138},
  pages = {479--488},
  issn = {0375-9687},
  doi = {10.1143/PTPS.138.479}
}

@article{lin.etal_2007,
  title = {Entanglement {{Entropy}} at {{Infinite-Randomness Fixed Points}} in {{Higher Dimensions}}},
  author = {Lin, Yu-Cheng and Igl{\'o}i, Ferenc and Rieger, Heiko},
  year = {2007},
  month = oct,
  journal = {Phys. Rev. Lett.},
  volume = {99},
  number = {14},
  pages = {147202},
  publisher = {{American Physical Society}},
  doi = {10.1103/PhysRevLett.99.147202}
}

@article{luo.etal_2022,
  title = {Postselected {{Entanglement}} between {{Two Atomic Ensembles Separated}} by 12.5 Km},
  author = {Luo, Xi-Yu and Yu, Yong and Liu, Jian-Long and Zheng, Ming-Yang and Wang, Chao-Yang and Wang, Bin and Li, Jun and Jiang, Xiao and Xie, Xiu-Ping and Zhang, Qiang and Bao, Xiao-Hui and Pan, Jian-Wei},
  year = {2022},
  month = jul,
  journal = {Phys. Rev. Lett.},
  volume = {129},
  number = {5},
  pages = {050503},
  publisher = {{American Physical Society}},
  doi = {10.1103/PhysRevLett.129.050503}
}

@article{ma.etal_1979,
  title = {Random {{Antiferromagnetic Chain}}},
  author = {Ma, Shang-Keng and Dasgupta, Chandan and Hu, Chin-Kun},
  year = {1979},
  month = nov,
  journal = {Phys. Rev. Lett.},
  volume = {43},
  number = {19},
  pages = {1434--1437},
  publisher = {{American Physical Society}},
  doi = {10.1103/PhysRevLett.43.1434}
}

@article{meignant.etal_2019,
  title = {Distributing Graph States over Arbitrary Quantum Networks},
  author = {Meignant, Cl{\'e}ment and Markham, Damian and Grosshans, Fr{\'e}d{\'e}ric},
  year = {2019},
  month = nov,
  journal = {Phys. Rev. A},
  volume = {100},
  number = {5},
  pages = {052333},
  publisher = {{American Physical Society}},
  doi = {10.1103/PhysRevA.100.052333}
}

@misc{meyer_,
  title = {Route {{Views}} \textendash{} {{University}} of {{Oregon Route Views Project}}},
  author = {Meyer, David},
  howpublished = {http://www.routeviews.org/routeviews/}
}

@article{motrunich.etal_2000,
  title = {Infinite-Randomness Quantum {{Ising}} Critical Fixed Points},
  author = {Motrunich, Olexei and Mau, Siun-Chuon and Huse, David A. and Fisher, Daniel S.},
  year = {2000},
  month = jan,
  journal = {Phys. Rev. B},
  volume = {61},
  number = {2},
  pages = {1160--1172},
  publisher = {{American Physical Society}},
  doi = {10.1103/PhysRevB.61.1160}
}

@article{peev.etal_2009,
  title = {The {{SECOQC}} Quantum Key Distribution Network in {{Vienna}}},
  author = {Peev, M. and Pacher, C. and All{\'e}aume, R. and Barreiro, C. and Bouda, J. and Boxleitner, W. and Debuisschert, T. and Diamanti, E. and Dianati, M. and Dynes, J. F. and Fasel, S. and Fossier, S. and F{\"u}rst, M. and Gautier, J.-D. and Gay, O. and Gisin, N. and Grangier, P. and Happe, A. and Hasani, Y. and Hentschel, M. and H{\"u}bel, H. and Humer, G. and L{\"a}nger, T. and Legr{\'e}, M. and Lieger, R. and Lodewyck, J. and Lor{\"u}nser, T. and L{\"u}tkenhaus, N. and Marhold, A. and Matyus, T. and Maurhart, O. and Monat, L. and Nauerth, S. and Page, J.-B. and Poppe, A. and Querasser, E. and Ribordy, G. and Robyr, S. and Salvail, L. and Sharpe, A. W. and Shields, A. J. and Stucki, D. and Suda, M. and Tamas, C. and Themel, T. and Thew, R. T. and Thoma, Y. and Treiber, A. and Trinkler, P. and {Tualle-Brouri}, R. and Vannel, F. and Walenta, N. and Weier, H. and Weinfurter, H. and Wimberger, I. and Yuan, Z. L. and Zbinden, H. and Zeilinger, A.},
  year = {2009},
  month = jul,
  journal = {New J. Phys.},
  volume = {11},
  number = {7},
  pages = {075001},
  publisher = {{IOP Publishing}},
  issn = {1367-2630},
  doi = {10.1088/1367-2630/11/7/075001},
  langid = {english}
}

@article{perseguers.etal_2008,
  title = {Entanglement Distribution in Pure-State Quantum Networks},
  author = {Perseguers, S{\'e}bastien and Cirac, J. Ignacio and Ac{\'i}n, Antonio and Lewenstein, Maciej and Wehr, Jan},
  year = {2008},
  month = feb,
  journal = {Phys. Rev. A},
  volume = {77},
  number = {2},
  pages = {022308},
  publisher = {{American Physical Society}},
  doi = {10.1103/PhysRevA.77.022308}
}

@article{pich.etal_1998,
  title = {Critical {{Behavior}} and {{Griffiths-McCoy Singularities}} in the {{Two-Dimensional Random Quantum Ising Ferromagnet}}},
  author = {Pich, C. and Young, A. P. and Rieger, H. and Kawashima, N.},
  year = {1998},
  month = dec,
  journal = {Phys. Rev. Lett.},
  volume = {81},
  number = {26},
  pages = {5916--5919},
  publisher = {{American Physical Society}},
  doi = {10.1103/PhysRevLett.81.5916}
}

@article{pirandola_2019a,
  title = {End-to-End Capacities of a Quantum Communication Network},
  author = {Pirandola, Stefano},
  year = {2019},
  month = may,
  journal = {Commun Phys},
  volume = {2},
  number = {1},
  pages = {1--10},
  publisher = {{Nature Publishing Group}},
  issn = {2399-3650},
  doi = {10.1038/s42005-019-0147-3},
  copyright = {2019 The Author(s)},
  langid = {english}
}

@article{rabbie.etal_2022,
  title = {Designing Quantum Networks Using Preexisting Infrastructure},
  author = {Rabbie, Julian and Chakraborty, Kaushik and Avis, Guus and Wehner, Stephanie},
  year = {2022},
  month = jan,
  journal = {npj Quantum Inf},
  volume = {8},
  number = {1},
  pages = {1--12},
  publisher = {{Nature Publishing Group}},
  issn = {2056-6387},
  doi = {10.1038/s41534-021-00501-3},
  copyright = {2022 The Author(s)},
  langid = {english}
}

@article{ravasz.barabasi_2003,
  title = {Hierarchical Organization in Complex Networks},
  author = {Ravasz, Erzs{\'e}bet and Barab{\'a}si, Albert-L{\'a}szl{\'o}},
  year = {2003},
  month = feb,
  journal = {Phys. Rev. E},
  volume = {67},
  number = {2},
  pages = {026112},
  publisher = {{American Physical Society}},
  doi = {10.1103/PhysRevE.67.026112}
}

@article{refael_moore04,
  title = {Entanglement {{Entropy}} of {{Random Quantum Critical Points}} in {{One Dimension}}},
  author = {Refael, G. and Moore, J. E.},
  year = {2004},
  month = dec,
  journal = {Phys. Rev. Lett.},
  volume = {93},
  number = {26},
  pages = {260602},
  publisher = {{American Physical Society}},
  doi = {10.1103/PhysRevLett.93.260602}
}

@article{ren.etal_2017,
  title = {Ground-to-Satellite Quantum Teleportation},
  author = {Ren, Ji-Gang and Xu, Ping and Yong, Hai-Lin and Zhang, Liang and Liao, Sheng-Kai and Yin, Juan and Liu, Wei-Yue and Cai, Wen-Qi and Yang, Meng and Li, Li and Yang, Kui-Xing and Han, Xuan and Yao, Yong-Qiang and Li, Ji and Wu, Hai-Yan and Wan, Song and Liu, Lei and Liu, Ding-Quan and Kuang, Yao-Wu and He, Zhi-Ping and Shang, Peng and Guo, Cheng and Zheng, Ru-Hua and Tian, Kai and Zhu, Zhen-Cai and Liu, Nai-Le and Lu, Chao-Yang and Shu, Rong and Chen, Yu-Ao and Peng, Cheng-Zhi and Wang, Jian-Yu and Pan, Jian-Wei},
  year = {2017},
  month = sep,
  journal = {Nature},
  volume = {549},
  number = {7670},
  pages = {70--73},
  publisher = {{Nature Publishing Group}},
  issn = {1476-4687},
  doi = {10.1038/nature23675},
  copyright = {2017 Macmillan Publishers Limited, part of Springer Nature. All rights reserved.},
  langid = {english}
}

@article{sachdev_2000,
  title = {Quantum {{Criticality}}: {{Competing Ground States}} in {{Low Dimensions}}},
  shorttitle = {Quantum {{Criticality}}},
  author = {Sachdev, Subir},
  year = {2000},
  month = apr,
  journal = {Science},
  volume = {288},
  number = {5465},
  pages = {475--480},
  publisher = {{American Association for the Advancement of Science}},
  doi = {10.1126/science.288.5465.475}
}

@book{sachdev_2011,
  title = {Quantum {{Phase Transitions}}},
  author = {Sachdev, Subir},
  year = {2011},
  edition = {Second},
  publisher = {{Cambridge University Press}},
  address = {{Cambridge}},
  doi = {10.1017/CBO9780511973765},
  isbn = {978-0-521-51468-2}
}

@article{sasaki.etal_2011,
  title = {Field Test of Quantum Key Distribution in the {{Tokyo QKD Network}}},
  author = {Sasaki, M. and Fujiwara, M. and Ishizuka, H. and Klaus, W. and Wakui, K. and Takeoka, M. and Miki, S. and Yamashita, T. and Wang, Z. and Tanaka, A. and Yoshino, K. and Nambu, Y. and Takahashi, S. and Tajima, A. and Tomita, A. and Domeki, T. and Hasegawa, T. and Sakai, Y. and Kobayashi, H. and Asai, T. and Shimizu, K. and Tokura, T. and Tsurumaru, T. and Matsui, M. and Honjo, T. and Tamaki, K. and Takesue, H. and Tokura, Y. and Dynes, J. F. and Dixon, A. R. and Sharpe, A. W. and Yuan, Z. L. and Shields, A. J. and Uchikoga, S. and Legr{\'e}, M. and Robyr, S. and Trinkler, P. and Monat, L. and Page, J.-B. and Ribordy, G. and Poppe, A. and Allacher, A. and Maurhart, O. and L{\"a}nger, T. and Peev, M. and Zeilinger, A.},
  year = {2011},
  month = may,
  journal = {Opt. Express, OE},
  volume = {19},
  number = {11},
  pages = {10387--10409},
  publisher = {{Optica Publishing Group}},
  issn = {1094-4087},
  doi = {10.1364/OE.19.010387},
  copyright = {\&\#169; 2011 OSA},
  langid = {english}
}

@article{satoh.etal_2012,
  title = {Quantum Network Coding for Quantum Repeaters},
  author = {Satoh, Takahiko and Le Gall, Fran{\c c}ois and Imai, Hiroshi},
  year = {2012},
  month = sep,
  journal = {Phys. Rev. A},
  volume = {86},
  number = {3},
  pages = {032331},
  publisher = {{American Physical Society}},
  doi = {10.1103/PhysRevA.86.032331}
}

@article{schoute.etal_2016,
  title = {Shortcuts to Quantum Network Routing},
  author = {Schoute, Eddie and Mancinska, Laura and Islam, Tanvirul and Kerenidis, Iordanis and Wehner, Stephanie},
  year = {2016},
  month = oct,
  journal = {ArXiv161005238 Quant-Ph},
  eprint = {1610.05238},
  eprinttype = {arxiv},
  primaryclass = {quant-ph},
  archiveprefix = {arXiv}
}

@article{senthil.majumdar_1996,
  title = {Critical {{Properties}} of {{Random Quantum Potts}} and {{Clock Models}}},
  author = {Senthil, T. and Majumdar, Satya N.},
  year = {1996},
  month = apr,
  journal = {Phys. Rev. Lett.},
  volume = {76},
  number = {16},
  pages = {3001--3004},
  publisher = {{American Physical Society}},
  doi = {10.1103/PhysRevLett.76.3001}
}

@article{senthil.sachdev_1996,
  title = {Higher {{Dimensional Realizations}} of {{Activated Dynamic Scaling}} at {{Random Quantum Transitions}}},
  author = {Senthil, T. and Sachdev, Subir},
  year = {1996},
  month = dec,
  journal = {Phys. Rev. Lett.},
  volume = {77},
  number = {26},
  pages = {5292--5295},
  publisher = {{American Physical Society}},
  doi = {10.1103/PhysRevLett.77.5292}
}

@inproceedings{shi.qian_2020,
  title = {Concurrent {{Entanglement Routing}} for {{Quantum Networks}}: {{Model}} and {{Designs}}},
  shorttitle = {Concurrent {{Entanglement Routing}} for {{Quantum Networks}}},
  booktitle = {Proc. {{Annu}}. {{Conf}}. {{ACM Spec}}. {{Interest Group Data Commun}}. {{Appl}}. {{Technol}}. {{Archit}}. {{Protoc}}. {{Comput}}. {{Commun}}.},
  author = {Shi, Shouqian and Qian, Chen},
  year = {2020},
  month = jul,
  series = {{{SIGCOMM}} '20},
  pages = {62--75},
  publisher = {{Association for Computing Machinery}},
  address = {{New York, NY, USA}},
  doi = {10.1145/3387514.3405853},
  isbn = {978-1-4503-7955-7}
}

@book{stauffer.aharony_2017,
  title = {Introduction {{To Percolation Theory}}: {{Second Edition}}},
  shorttitle = {Introduction {{To Percolation Theory}}},
  author = {Stauffer, Dietrich and Aharony, Ammon},
  year = {2017},
  month = jan,
  edition = {Second},
  publisher = {{Taylor \& Francis}},
  address = {{London}},
  doi = {10.1201/9781315274386},
  isbn = {978-1-315-27438-6}
}

@article{stinchcombe_1981,
  title = {Diluted Quantum Transverse {{Ising}} Model},
  author = {Stinchcombe, R. B.},
  year = {1981},
  month = apr,
  journal = {J. Phys. C: Solid State Phys.},
  volume = {14},
  number = {10},
  pages = {L263--L267},
  publisher = {{IOP Publishing}},
  issn = {0022-3719},
  doi = {10.1088/0022-3719/14/10/003},
  langid = {english}
}

@article{sundar.etal_2021,
  title = {Response of Quantum Spin Networks to Attacks},
  author = {Sundar, Bhuvanesh and Walschaers, Mattia and Parigi, Valentina and Carr, Lincoln D.},
  year = {2021},
  month = may,
  journal = {J. Phys. Complex.},
  volume = {2},
  number = {3},
  pages = {035008},
  publisher = {{IOP Publishing}},
  issn = {2632-072X},
  doi = {10.1088/2632-072X/abf5c2},
  langid = {english}
}

@article{szalay,
  title = {Multipartite Entanglement Measures},
  author = {family=Szalay, given=Szil{\'a}rd, given-i={Sz}},
  year = {2015},
  month = oct,
  journal = {Phys. Rev. A},
  volume = {92},
  number = {4},
  pages = {042329},
  publisher = {{American Physical Society}},
  doi = {10.1103/PhysRevA.92.042329}
}

@article{vanleent.etal_2022,
  title = {Entangling Single Atoms over 33 Km Telecom Fibre},
  author = {{van Leent}, Tim and Bock, Matthias and Fertig, Florian and Garthoff, Robert and Eppelt, Sebastian and Zhou, Yiru and Malik, Pooja and Seubert, Matthias and Bauer, Tobias and Rosenfeld, Wenjamin and Zhang, Wei and Becher, Christoph and Weinfurter, Harald},
  year = {2022},
  month = jul,
  journal = {Nature},
  volume = {607},
  number = {7917},
  pages = {69--73},
  publisher = {{Nature Publishing Group}},
  issn = {1476-4687},
  doi = {10.1038/s41586-022-04764-4},
  copyright = {2022 The Author(s)},
  langid = {english}
}

@article{vidal,
  title = {Entanglement in {{Quantum Critical Phenomena}}},
  author = {Vidal, G. and Latorre, J. I. and Rico, E. and Kitaev, A.},
  year = {2003},
  month = jun,
  journal = {Phys. Rev. Lett.},
  volume = {90},
  number = {22},
  pages = {227902},
  publisher = {{American Physical Society}},
  doi = {10.1103/PhysRevLett.90.227902}
}

@inproceedings{wang.etal_2006,
  title = {Visualization of Large Hierarchical Data by Circle Packing},
  booktitle = {Proc. {{SIGCHI Conf}}. {{Hum}}. {{Factors Comput}}. {{Syst}}.},
  author = {Wang, Weixin and Wang, Hui and Dai, Guozhong and Wang, Hongan},
  year = {2006},
  month = apr,
  series = {{{CHI}} '06},
  pages = {517--520},
  publisher = {{Association for Computing Machinery}},
  address = {{New York, NY, USA}},
  doi = {10.1145/1124772.1124851},
  isbn = {978-1-59593-372-0}
}

@article{wehner.etal_2018,
  title = {Quantum Internet: {{A}} Vision for the Road Ahead},
  shorttitle = {Quantum Internet},
  author = {Wehner, Stephanie and Elkouss, David and Hanson, Ronald},
  year = {2018},
  month = oct,
  journal = {Science},
  volume = {362},
  number = {6412},
  publisher = {{American Association for the Advancement of Science}},
  issn = {0036-8075, 1095-9203},
  doi = {10.1126/science.aam9288},
  chapter = {Review},
  copyright = {Copyright \textcopyright{} 2018 The Authors, some rights reserved; exclusive licensee American Association for the Advancement of Science. No claim to original U.S. Government Works. http://www.sciencemag.org/about/science-licenses-journal-article-reuseThis is an article distributed under the terms of the Science Journals Default License.},
  langid = {english},
  pmid = {30337383}
}

@article{wei.etal_2022,
  title = {Towards Real-World Quantum Networks: A Review},
  shorttitle = {Towards Real-World Quantum Networks},
  author = {Wei, Shi-Hai and Jing, Bo and Zhang, Xue-Ying and Liao, Jin-Yu and Yuan, Chen-Zhi and Fan, Bo-Yu and Lyu, Chen and Zhou, Dian-Li and Wang, You and Deng, Guang-Wei and Song, Hai-Zhi and Oblak, Daniel and Guo, Guang-Can and Zhou, Qiang},
  year = {2022},
  month = jan,
  journal = {ArXiv220104802 Quant-Ph},
  eprint = {2201.04802},
  eprinttype = {arxiv},
  primaryclass = {quant-ph},
  archiveprefix = {arXiv}
}

@article{yin.etal_2017,
  title = {Satellite-Based Entanglement Distribution over 1200 Kilometers},
  author = {Yin, Juan and Cao, Yuan and Li, Yu-Huai and Liao, Sheng-Kai and Zhang, Liang and Ren, Ji-Gang and Cai, Wen-Qi and Liu, Wei-Yue and Li, Bo and Dai, Hui and Li, Guang-Bing and Lu, Qi-Ming and Gong, Yun-Hong and Xu, Yu and Li, Shuang-Lin and Li, Feng-Zhi and Yin, Ya-Yun and Jiang, Zi-Qing and Li, Ming and Jia, Jian-Jun and Ren, Ge and He, Dong and Zhou, Yi-Lin and Zhang, Xiao-Xiang and Wang, Na and Chang, Xiang and Zhu, Zhen-Cai and Liu, Nai-Le and Chen, Yu-Ao and Lu, Chao-Yang and Shu, Rong and Peng, Cheng-Zhi and Wang, Jian-Yu and Pan, Jian-Wei},
  year = {2017},
  month = jun,
  journal = {Science},
  volume = {356},
  number = {6343},
  pages = {1140--1144},
  publisher = {{American Association for the Advancement of Science}},
  issn = {0036-8075, 1095-9203},
  doi = {10.1126/science.aan3211},
  chapter = {Research Articles},
  copyright = {Copyright \textcopyright{} 2017 The Authors, some rights reserved; exclusive licensee American Association for the Advancement of Science. No claim to original U.S. Government Works. http://www.sciencemag.org/about/science-licenses-journal-article-reuseThis is an article distributed under the terms of the Science Journals Default License.},
  langid = {english},
  pmid = {28619937}
}

@article{yu.etal_2008,
  title = {Entanglement Entropy in the Two-Dimensional Random Transverse Field {{Ising}} Model},
  author = {Yu, Rong and Saleur, Hubert and Haas, Stephan},
  year = {2008},
  month = apr,
  journal = {Phys. Rev. B},
  volume = {77},
  number = {14},
  pages = {140402},
  publisher = {{American Physical Society}},
  doi = {10.1103/PhysRevB.77.140402}
}

@article{zhang.zhuang_2021,
  title = {Quantum Internet under Random Breakdowns and Intentional Attacks},
  author = {Zhang, Bingzhi and Zhuang, Quntao},
  year = {2021},
  month = jul,
  journal = {Quantum Sci. Technol.},
  volume = {6},
  number = {4},
  pages = {045007},
  publisher = {{IOP Publishing}},
  issn = {2058-9565},
  doi = {10.1088/2058-9565/ac1041},
  langid = {english}
}

@article{zhuang.zhang_2021,
  title = {Quantum Communication Capacity Transition of Complex Quantum Networks},
  author = {Zhuang, Quntao and Zhang, Bingzhi},
  year = {2021},
  month = aug,
  journal = {Phys. Rev. A},
  volume = {104},
  number = {2},
  pages = {022608},
  publisher = {{American Physical Society}},
  doi = {10.1103/PhysRevA.104.022608}
}

@article{zou.etal_2022,
  title = {Multipartite Entanglement in the Random {{Ising}} Chain},
  author = {Zou, Jay S. and Ansell, Helen S. and Kov{\'a}cs, Istv{\'a}n A.},
  year = {2022},
  month = aug,
  journal = {Phys. Rev. B},
  volume = {106},
  number = {5},
  pages = {054201},
  publisher = {{American Physical Society}},
  doi = {10.1103/PhysRevB.106.054201}
}

\end{document}